\documentclass[a4paper,11pt]{article}
\usepackage[left=2.15cm,right=2.15cm,top=2.15cm,
bottom=2.15cm]{geometry}
\usepackage[usenames,dvipsnames,svgnames,table]{xcolor}
\usepackage{psfrag,amsmath,amsfonts,blkarray,amssymb,latexsym,amsthm,verbatim,color,lscape,multirow,graphicx,subcaption}
\usepackage[figuresright]{rotating}
\usepackage[compress]{natbib}
\usepackage{url}
\usepackage{hyperref}
\hypersetup{colorlinks,
citecolor=black,
filecolor=black,
linkcolor=black,
urlcolor=black,
pdftex}
\usepackage{pdfpages}
\usepackage{booktabs}

\usepackage{tikzducks}

\usepackage{colortbl}
\definecolor{gray}{rgb}{0.9,0.9,0.9}

\usepackage{times}
\usepackage{blindtext}
\usepackage[font={footnotesize,it}]{caption}

\usepackage{sectsty}  
\subsectionfont{\normalsize\bf}
\sectionfont{\large\bf}

\usepackage[symbol]{footmisc}

\usepackage{algorithm}
\usepackage[noend]{algpseudocode}

\usepackage{tabularx}
\newcommand\setrow[1]{\gdef\rowmac{#1}#1\ignorespaces}
\newcommand\clearrow{\global\let\rowmac\relax}
\clearrow

\let\proglang=\textsf
\newcommand{\pkg}[1]{{\fontseries{b}\selectfont #1}}

\def\R{\mathbf{R}}
\def\S{\mathbf{S}}
\def\y{\mathbf{y}}

\def\x{\mathbf{x}}

\def\s{\sigma}

\def\u{\mathbf{u}}
\def\T{\mathbf{T}}

\def\X{\mathbf{X}}

\def\bfpi{\boldsymbol{\pi}}

\def\g{\gamma}

\def\de{\delta}
\def\debf{\boldsymbol{\delta}}
\def\gbf{\boldsymbol{\gamma}}

\def\th{\theta}

\def\thbf{\boldsymbol{\theta}}

\def\pibf{\boldsymbol{\pi}}

\def\Sbf{\boldsymbol{\Sigma}}
\def\sbf{\boldsymbol{\sigma}}
\def\pibf{\boldsymbol{\pi}}

\def \blue#1{\textcolor{black}{#1}}

\def\bmu{\boldsymbol{\mu}}

\begin{document}

\title{1-truncated C-vine copula mixed models for network meta-analysis of multiple diagnostic tests}

\author{Aristidis~K.~Nikoloulopoulos \footnote{\href{mailto:a.nikoloulopoulos@uea.ac.uk}{a.nikoloulopoulos@uea.ac.uk},  School of Engineering, Mathematics and Physics, University of East Anglia, Norwich NR47TJ, U.K.} }

\date{}

\maketitle

\baselineskip=24pt

\begin{abstract}
\baselineskip=24pt
\noindent  As meta-analysis of multiple diagnostic tests impacts clinical decision making and patient health, there is growing interest in statistical models that synthesize evidence from studies comparing multiple diagnostic tests. To compare the accuracy of multiple diagnostic tests in a single study, three designs are commonly used: (i) the multiple test comparison design; (ii) the randomized design, and (iii) the non-comparative design. Generalized linear mixed models (GLMMs) are currently the recommended approach for jointly meta-analyzing data from all three designs, enabling simultaneous inference. In this context, 1-truncated C-vine copula mixed models are proposed as a flexible and powerful alternative. These models generalize the GLMM framework by allowing for arbitrary univariate distributions of the random effects and capturing tail dependencies and asymmetries.  We demonstrate the utility of our methods with an extensive simulation study and by insightfully re-analysing   a case study on the network meta-analysis of diagnostic tests for deep vein thrombosis. Findings indicate that 1-truncated C-vine copula mixed models can offer improvements over GLMMs, supporting their adoption for network meta-analysis of multiple diagnostic tests.


\end{abstract}

\section{Introduction}

Comparative effectiveness research relies on accurate evaluation of clinical outcomes. With the proliferation of diagnostic tools and rising costs, there is growing demand for rigorous comparisons of multiple diagnostic tests in clinical practice. Advances in disease understanding and technology have led to the development of numerous diagnostic options, making evidence-based selection essential.

To assess the accuracy of multiple diagnostic tests within a single study, three common designs are typically employed \citep{Takwoingi-etal-2013}: (i) the multiple test comparison design, in which all participants receive every candidate test and are verified against a gold standard; (ii) the randomized design, where participants are randomly assigned to one candidate test, with all subsequently verified by a gold standard; and (iii) the non-comparative design, where distinct groups of participants are used to compare a candidate test either with a gold standard or with another candidate test.

Meta-analysis methods have been widely adopted to improve the estimation of diagnostic test accuracy by synthesizing evidence across studies (e.g., \citealt{JacksonRileyWhite2011}). 
Meta-analysis of diagnostic tests helps identify the most accurate test or combination of tests for a given condition, guiding clinicians toward informed decisions that improve patient outcomes while reducing costs \citep{Chartrand2012}. For instance, meta-analyses of COVID-19 diagnostic studies have clarified which tests are most reliable across settings and populations, enabling timely and accurate diagnoses that directly impact treatment and patient health \citep{Vandenberg2021}.
The rapid evolution of diagnostic strategies and the diversity of study designs create several challenges for comparing multiple tests. 
Diagnostic accuracy estimates are frequently associated because many studies assess multiple diagnostic tests within the same population, and failure to account for these associations can lead to a loss of efficiency. In addition, most meta-analyses are based on studies that evaluate only a subset of the tests of interest, which complicates direct comparisons across tests. Variation in reference standards, some using a gold standard and others relying on error-prone alternatives, further introduces bias. These challenges are compounded by differences in study design and by varying degrees of between-study heterogeneity across tests, making robust meta-analysis and interpretation more challenging.

Given the diversity of study designs, a flexible meta-analytic framework is needed to integrate evidence from all three approaches and enable effective ranking of candidate tests. 
Nevertheless, the  majority of  available meta-analysis methods for comparing
multiple tests (\citealt{trikalinos-etal-2014-rsm}; \citealt{dimou-etal2016};
 \citealt{hoyer&kuss-2016-smmr};
 \citealt{Nyaga-etal-2018n}; \citealt{Nyaga-etal-2018b}; \citealt{nikoloulopoulos-2018-smmr,
Nikoloulopoulos2020-factorREMADA,Nikoloulopoulos-2024}) focuses  either to a multiple tests design or requires
each patient's true disease status to be known, that is, that each
study's reference test is a gold standard

To the best of our knowledge,
there only exist three meta-analysis methods \citep{Menten&lessaffre-2015,Ma-etal-2018-biostatistics,Lian-eta-al-2019-jasa} for comparing multiple
diagnostic tests that can simultaneously incorporate studies with
different designs and studies with or without a gold standard
test. 
\cite{Menten&lessaffre-2015}  introduced a contrast-based model, where  the transformed latent  sensitivity and specificity of each test  is a linear combination of an overall  latent probability probability of testing positive on alls tests and 
contrasts of the latent log odds ratios for sensitivity and specificity of the $K-1$ tests compared to a baseline test. The latent log odds ratios are assumed to follow the multivariate normal (MVN) distribution.  The variance-covariance matrix models the variances and co-variances of the contrasts and contributes in a complex manner to
the variances and co-variances of the transformed  latent  sensitivity and specificity of each test.  
Furthermore, the model identification becomes difficult as the number of tests included increases.
To enable model identification, the authors recommended simplified variance-covariance structures as a diagonal or block diagonal variance
covariance matrix. However, these simplifications    do not properly account for correlations between the diagnostic  tests. 
\cite{Ma-etal-2018-biostatistics} and \cite{Lian-eta-al-2019-jasa}  properly accounted for   correlations between multiple tests. 
Nevertheless, they also relied to the MVN distribution  of transformed latent proportions which has restricted properties, such as a linear correlation structure and  normal margins. These limitations  might lead to biased meta-analytic estimates of diagnostic test accuracy.

In order to create a flexible distribution to model  the latent prevalence, sensitivities and specificities, we exploit the use of  
regular vine copulas \citep{Bedford&Cooke02,Erhardt-etal-2015-biometrics}.  
We use their boundary case, namely   a C-vine copula.
 C-vine copulas  have become important in  many applications areas such as finance
\citep{aasetal09,nikoloulopoulos&joe&li11,krupskii-joe-2013}, psychometrics \citep{nikoloulopoulos&joe12,Kadhem&Nikoloulopoulos-2021}, and biological sciences
\citep{Nikoloulopoulos2015c,Erhardt&Czado2018-JRSSC}, to  name just a few,  in order to deal with dependence in the joint tails. 
Another boundary case of regular vine copulas is the D-vine copula, but this parametric family of copulas is suitable if there is serial-dependence among the variables  (e.g., \citealt{Barthel-etal-2019-biom,Hoque&Acar&Torabi-2022-Biometrics}), which apparently is not the case in this application area. 
We propose a C-vine  
copula mixed model  as an extension of the  model in  \cite{Ma-etal-2018-biostatistics} by  using a C-vine copula   representation of the random effects distribution with both normal and beta margins.  
The proposed model (a) includes the model in  \cite{Ma-etal-2018-biostatistics}  as a special case, and  thus can  integrate evidence from studies of all three designs,  can pool results across both comparative and non-comparative studies,   and can compare any candidate test either with a gold standard or with another imperfect candidate test, (b) can have arbitrary univariate distributions for the random effects and (c) can  provide tail dependencies or asymmetries.

The remainder of the paper proceeds as follows. Section \ref{model}  introduces the C-vine copula mixed model for network meta-analysis of multiple diagnostic tests, provides computational details for maximum likelihood (ML) estimation, and discusses its relationship with the model in \cite{Ma-etal-2018-biostatistics}.
Section \ref{miss-section}  gauges the small-sample efficiency and robustness of the ML estimation of the   proposed C-vine copula  mixed model.   
Section \ref{app-section} insightfully re-analyzes     a case study on the network meta-analysis of diagnostic tests for deep vein thrombosis. 
We conclude with some discussion in Section \ref{discussion},  followed by a brief section with software details.

\section{\label{model} The 1-truncated C-vine copula mixed model}
In this section, we introduce the 1-truncated 
C-vine copula mixed model  for network meta-analysis of multiple diagnostic tests. As in \cite{Ma-etal-2018-biostatistics} and \cite{Lian-eta-al-2019-jasa}, we treat studies with randomized and non-comparative designs as if they followed a multiple test comparison framework, in which all subjects across all studies were assessed using every candidate diagnostic test as well as a gold-standard test. In practice, however, most studies evaluate only a subset of the full set of tests of interest, and outcomes from tests not included in a given study are regarded as missing data. By jointly comparing all candidate tests with the gold standard, the proposed approach leverages all available information and enables information sharing across studies.

\subsection{Within-study model}
The within-study model is the same as in \cite{Ma-etal-2018-biostatistics} and \cite{Lian-eta-al-2019-jasa}.  
In what follows we adopt the notation of \cite{Ma-etal-2018-biostatistics}.  
Let $\T = \{T_0, T_1, \dots, T_K\}$ be a set of $K+1$ binary diagnostic tests, where $T_0$ denotes a gold standard and $T_1, \dots, T_K$ stand for candidate tests under evaluation. Suppose we have a collection of $i = 1, \dots, N$ studies, where each of them reports outcomes of tests in a subset of $\T$. In the $i$th study, for $k = 0, 1, \dots, K$, let $y_{ijk}$ be the test outcome of $T_k$ on subject $j$ ($y_{ijk} = 1$ if positive and $0$ if negative) and let 
$t_{ij0}$ be the indicator if  the $j$th subject is tested by the gold standard ($t_{ij0} = 1$ if tested  and $0$ if not).
Let $X$ be the transformed latent disease prevalence: $X= l\bigl[\Pr(y_{ij0} = 1)\bigr], \; i = 1, \dots, N$, $l(p)$ is a link function as the commonly used $\mbox{logit}(p)=\log(\frac{p}{1-p})$. For $k = 1, \dots, K$, let $X_{1k}$ and $X_{0k}$ denote the transformed  latent sensitivity and specificity for the $k$th test, respectively: $X_{1k} = l\bigl[\Pr(y_{ijk} = 1 | y_{ij0} = 1)\bigr]$ and $X_{0k}= l\bigl[\Pr(y_{ijk} = 0 | y_{ij0} = 0)\bigr]$. Denote $K_{ij}$ as the set of candidate tests conducted on subject $j$ ($j = 1, \dots, J_i$) in the $i$th study, $\y_{ij} = \{y_{ijk} : k \in K_{ij}\}$ as the collection of candidate test outcomes for this subject, and $\X_{d}=(X_{d1},\ldots,X_{dK})$ the vector of  of candidate test sensitivities ($d=1$) or specificities ($d=0$).
In contrast with \cite{Ma-etal-2018-biostatistics}, we prefer to use notation that distinguishes
unobserved  from observed variables. For observed variables we use $i$ since they are observed per individual study $i$, but this is not the case for the latent variables or random effects.

Let $x$ and  $\x_d=(x_{d1},\ldots,x_{dK}),\,d=0,1$ be realizations of the latent random variable $X$ and  vector $\X_d$, respectively.
To derive the likelihood for the $j$th subject in the $i$th study, 
we first consider a subject that is tested by the gold standard test ($t_{ij0} = 1$) such that the true disease status is known. Conditional independence is assumed such that candidate test results are independent given the disease status. 
The probability of the test outcomes for a diseased subject, given the random effects, is calculated as:
\begin{eqnarray}\label{d1}
\Pr\Bigl(\y_{ij}, y_{ij0} = 1 \mid (X=x,\X_{1}=\x_{1})\Bigr) &=& P(y_{ij0} = 1)\Pr\Bigl(\y_{ij}\mid (y_{ij0}=1,X=x,\X_{1}=\x_{1})\Bigr)\nonumber\\
&=&
l^{-1}(x) \prod_{k \in K_{ij}} \bigl[l^{-1}(x_{1k})\bigr]^{y_{ijk}} \bigl[1-l^{-1}(x_{1k})\bigr]^{1-y_{ijk}}\nonumber\\
&:=&g_1(\y_{ij}; x,\x_{1}).
\end{eqnarray}

\noindent Similarly, the probability for a non-diseased subject $j$ in study $i$ is given by:
\begin{eqnarray}\label{d0}
\Pr\Bigl(\y_{ij}, y_{ij0} = 0 \mid (X=x,\X_{0}=\x_{0})\Bigr) &=& \Pr(y_{ij0} = 0)\Pr(\y_{ij} \mid (y_{ij0}=0,X=x,\X_{0}=\x_{0})\Bigr)\nonumber\\&=&\bigl[1 - l^{-1}(x)\bigr] \prod_{k \in K_{ij}} \bigl[l^{-1}(x_{0k})\bigr]^{1-y_{ijk}} \bigl[1 - l^{-1}(x_{0k})\bigr]^{y_{ijk}}\nonumber\\&:=&g_0(\y_{ij}; x,\x_{0}).
\end{eqnarray}

\noindent Now consider the setting where the subject $j$ has not been tested by the gold standard $T_0$ (i.e., $t_{ij0} = 0$). By the law of total probability, the probability of the test outcomes $y_{ij}$, given the random effects, is given by the sum of the probabilities in (\ref{d1}) and (\ref{d0}).

In general, the probability of test outcomes for subject $j$ (with or without the gold standard) can be written in the following unified form:
\begin{multline}
\label{within}
g(\y_{ij},y_{ij0},t_{ij0};x,\x_{1},\x_{0})=\\\Bigl[g_1(\y_{ij}; x,\x_{1})\Bigr]^{t_{ij0} y_{ij0}} \Bigl[g_0(\y_{ij}; x,\x_{0})\Bigr]^{t_{ij0}(1 - y_{ij0})} 
\Bigl[g_1(\y_{ij}; x,\x_{1})+g_0(\y_{ij}; x,\x_{0})\Bigr]^{1 - t_{ij0}}.
\end{multline}

\subsection{Between studies model}

 For the between studies model,
there is a common latent variable $X$ and  different latent variables $(X_{1k},X_{0k})$
for each test $k$, but they are dependent. 
To model the dependence among the latent variables $X,X_{dk},\,d=0,1,\,k=1,\ldots,K$ we employ copulas. A copula is a multivariate cumulative distribution function (cdf) with uniform $U(0,1)$ margins \citep{joe97,joe2014}. The power of copulas for dependence modelling is due to the dependence structure being considered separate from the univariate margins; see, for example, Section 1.6 of \cite{joe97}. 
Denote the univariate  cdf of $X$ by $F\bigl(\cdot;l(\pi),\de\bigr)$, where $\pi$ is the meta-analytic parameter of disease prevalence and  $\de$ is the between-study variability for disease prevalence.  
Similarly, for $d=0,1,\,k=1,\ldots,K$ denote the univariate  cdf of $X_{dk}$ by $F\bigl(\cdot;l(\pi_{dk}),\de_{dk}\bigr)$, where $\pi_{dk}$ is the meta-analytic parameter of sensitivity ($d=1$) or  specificity ($d=0$) for test $k$,  $\de_{dk}$ is the between-study variability for  sensitivity  ($d=1$) or specificity ($d=0$) for test $d$. The choices of the  $F\bigl(\cdot;l(\pi),\de\bigr)$ and  $l$ are given in Table \ref{choices}.  If the Beta$(\pi,\gamma)$ distribution is used for the marginal modelling of
the latent proportions, then  
 one  does not have to transform the prevalence, sensitivities and specificities and can work on the original scale.

 \begin{table}[!h]
\begin{center}
\caption{\label{choices}The choices of the  $F\bigl(\cdot;l(\pi),\de\bigr)$ and  $l$ in the 1-truncated C-vine copula mixed model.}
 \begin{tabular*}{\linewidth}{@{\extracolsep{\fill}}cccc}\toprule
 $F\bigl(\cdot;l(\pi),\de\bigr)$ & $l$ & $\pi$ & $\de$\\\midrule
$N(\mu,\s)$ & logit, probit, cloglog & $l^{-1}(\mu)$&$\s$\\
Beta$(\pi,\gamma)$ & identity & $\pi$ & $\gamma$\\
\bottomrule
\end{tabular*}
\end{center}
\end{table}

In multivariate models with copulas, a copula or multivariate uniform distribution is combined with a set of univariate margins. That is, if a  $(2K+1)$-dimensional  parametric
family of copulas $C(\cdot;\thbf)$ is combined with the  
 parametric models $F\bigl(\cdot;l(\pi),\de\bigr)$ and $F\bigl(\cdot;l(\pi_{dk}),\de_{dk}\bigr),\,d=0,1,\,k=1,\ldots,K$, then $
 C\Bigl(F\bigl(x;l(\pi),\de\bigr),F\bigl(x_{11}; l(\pi_{11}),\de_{11}\bigr),\ldots,F\bigl(x_{1K};l(\pi_{1K}),\de_{1K}\bigr),\ldots,F\bigl(x_{01};l(\pi_{01}),$ $\de_{01}\bigr),\ldots,F\bigl(x_{0K};l(\pi_{0K}),\de_{0K}\bigr);\thbf\Bigr)$
is a multivariate parametric model with univariate margins $F\bigl(\cdot;$ $l(\pi),\de\bigr)$  and $F\bigl(\cdot;l(\pi_{dk}),\de_{dk}\bigr),\,d=0,1,\,k=1,\ldots,K$.
This is equivalent to assuming that the latent variables $X$ and $X_{dk},\,d=0,1,\,k=1,\ldots,K$ have been transformed to standard uniform latent variables $U=F\bigl(X;l(\pi),\de\bigr)$ and $U_{dk}=F\bigl(X_{dk};l(\pi_{dk}),\de_{dk}\bigr),\,d=0,1,\,k=1,\ldots,K$. 
So we assume that $(U,U_{11},\ldots,U_{1K},U_{01},\ldots,U_{0K})$ is a random vector with $U,U_{kt}\sim U(0,1)$ and  joint cdf  given by $C (u,u_{11},\ldots,u_{1K},u_{01},\ldots,u_{0K};\thbf)$,
where $C$ is a $(2K+1)$-dimensional C-vine copula, which is built via successive mixing from $2K$ bivariate marginal copulas at level 1 and $K(2K-1)$
bivariate conditional copulas at higher levels \citep{nikoloulopoulos&joe&li11}.

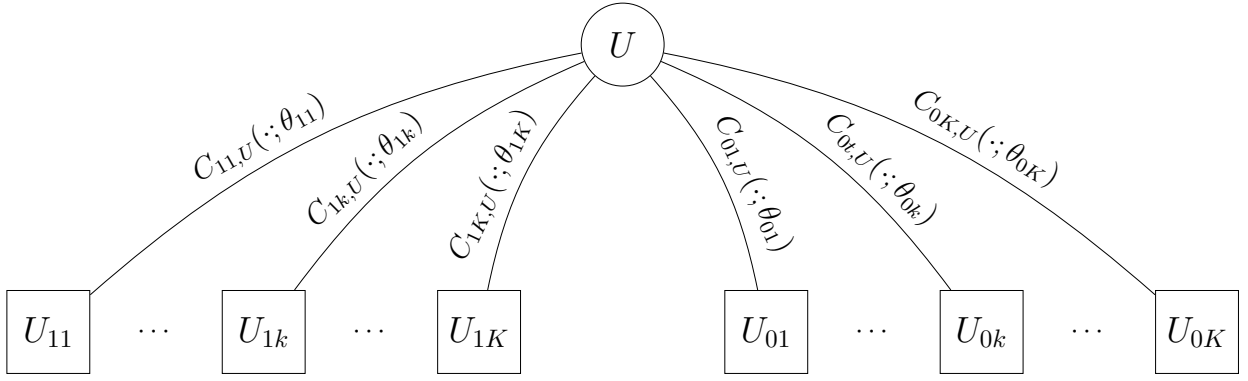
\begin{figure}[h!]
\scalebox{0.95}{
    \begin{tikzpicture}
        [square/.style={
            draw,
            fill=white!,
            minimum width=3em,
            minimum height=3em,
            node contents={#1}}
            ]

        \node at (0,1) [circle,draw,fill=white!,minimum width=3em] (X0) {\Large $U$};

	    \node at (-8,-3) (Y11) [square={\Large $U_{11}$}]; 
         		\node at  (-6.5,-3)  {$\cdots$}; 
        \node at (-5,-3) (Yj1) [square={\Large $U_{1k}$}]; 
                  \node at  (-3.5,-3)  {$\cdots$}; 
        \node at (-2,-3) (Yd11) [square={\Large $U_{1K}$}]; 

\path(X0)edge[black,bend right=15] node[above,rotate=30, midway,pos=0.60]{\large $C_{11,U}(
\cdot;\theta_{11})$}  (Y11);
\path(X0)edge[black,bend right=15] node[above,rotate=45, midway,pos=0.65]{\large $C_{1k,U}(
\cdot;\theta_{1k})$} (Yj1);
\path(X0)edge[black,bend right=15] node[above,rotate=65, midway,pos=0.60]{\large $C_{1K,U}(
\cdot;\theta_{1K})$} (Yd11);

        \node at (2,-3) (Y1G) [square={\Large $U_{01}$}];
         		\node at  (3.5,-3)  {$\cdots$};
        \node at (5,-3) (YjG) [square={\Large $U_{0k}$}];
                  \node at  (6.5,-3)  {$\cdots$};
        \node at (8,-3) (YdGG) [square={\Large $U_{0K}$}];

\path(X0) edge[black,bend right=-15] node[above,rotate=-65, midway,pos= 0.60]{\large $C_{01,U}(
\cdot;\theta_{01})$} (Y1G);
\path(X0) edge[black,bend right=-15] node[above,rotate=-45, midway,pos=0.65]{\large $C_{0t,U}(
\cdot;\theta_{0k})$} (YjG);
\path(X0) edge[black,bend right=-15] node[above,rotate=-30, midway,pos=0.60]{\large $C_{0K,U}(
\cdot;\theta_{0K})$} (YdGG);

    \end{tikzpicture}
}
\caption{\label{graphical-representation}Graphical representation of the 1-truncated  C-vine copula model which consists on $2K+1$ nodes with the latent variables  $U=F(X), U_{dk}=F(X_{dk}),\,d=0,1,\,k=1,\ldots,K$  and $2K$ edges.  Each edge is allied  with a  copula $C_{dk,U}(;\th_{dk})$ associated with the latent pair $(X,X_{dk})$ with $\th_{dk}$ being the bivariate copula parameter  denoting the association between  $X_{dk}$, corresponding to the latent sensitivity ($d=1$) and latent specificity ($d=0$) for each test $k$, and the latent disease prevalence  $X$.}
\end{figure}

For parsimony, we use a 1-truncated C-vine copula which $2K$  parametric bivariate copulas at level 1 and independence copulas at higher levels. 
For the 1-truncated C-vine copula, the pairs at level 1 are $U,U_{dk}$, for
$d=0,1,\,k=1,\ldots,K$, 
and for  higher levels  the (conditional)
copula pairs are set to independence. 
That is 
the 1-truncated C-vine copula has $2K$ bivariate copulas $C_{dk,U}(\cdot;\th_{dk})$ that link $U_{dk},\,d=0,1,\,k=1,\ldots,K$ with \blue{the common latent variable} $U$  in the 1st level  and independence copulas in all the remaining levels  (truncated after the 1st level). \blue{That is the common  latent variable $U$  drives the dependence between the  latent variables $U_{dk},\,d=0,1,\,k=1,\ldots,K$.}
Figure \ref{graphical-representation} depicts the graphical representation of the 1-truncated C-vine copula model. 
\cite{joeetal10} 
\blue{have shown that a vine copula displays (tail) dependence in all bivariate margins provided that the pair-copulas in the first level possess (tail) dependence; higher-level pair-copulas may be independence copulas without loss of overall (tail) dependence. This insight justifies truncating the vine after the first level, creating a parsimonious model that retains the essential dependence structure.}
Hence, the (tail) dependence between the \blue{common} latent variable $U$ and each of the  latent variables $U_{dk},\,d=0,1,\,k=1,\ldots,K$ is inherited to the (tail) dependence between the latent variables $U_{dk},\,d=0,1,\,k=1,\ldots,K$, i.e., between the   latent variables $X_{dk},\,d=0,1,\,k=1,\ldots,K$, corresponding to the latent sensitivity ($d=1$) and latent specificity ($d=0$) for  test $k$, through the dependence invariance property of copulas. That provides 
the  theoretical justification for the idea to model the  (tail) dependence in the first level and then just use the independence copulas to model conditional dependence at higher levels without sacrificing the (tail) dependence of the vine copula distribution.

To this end, the stochastic representation of the between studies model takes the form
\begin{eqnarray}\label{between}
&&\Bigl(F\bigl(X;l(\pi),\de\bigr),F\bigl(X_{11};l(\pi_{11}),\de_{11}\bigr),\ldots,
F\bigl(X_{1k};l(\pi_{1k}),\de_{1k}\bigr),\ldots,F\bigl(X_{1K};l(\pi_{1K}),\de_{1K}\bigr),\nonumber\\
&&F\bigl(X_{01};l(\pi_{01}),\de_{01}\bigr),\ldots,
F\bigl(X_{0k};l(\pi_{0k}),\de_{0k}\bigr),\ldots,F\bigl(X_{0K};l(\pi_{0K}),\de_{0K}\bigr)\Bigr)\sim C(\cdot;\thbf),
\end{eqnarray} 
where $\thbf=(\th_{11},\ldots,\th_{1k},\ldots,\th_{1K},\th_{01},\ldots,\th_{0k},\ldots\th_{0K})$.

The models in (\ref{within}) and (\ref{between}) together specify a  1-truncated C-vine copula mixed  model with joint likelihood
\begin{multline}\label{mixed-cop-likelihood}
L(\pi,\pi_{11},\ldots,\pi_{1K},\pi_{0K},\ldots,\pi_{0K},\de,\de_{11},\ldots,\de_{1K},\de_{01},\ldots,\de_{0K},\thbf)=\\
\prod_{i=1}^N\int_{[0,1]^{2K+1}}
\prod_{j=1}^{J_i}g\Bigl(\y_{ij},y_{ij0},t_{ij0}\mid x,\x_{1},\x_{0}\Bigr)dC(u,u_{11},\ldots, u_{1K},u_{01},\ldots,u_{0K};\thbf),
\end{multline}
where   $x=F^{-1}\bigl(u;l(\pi),\de\bigr)$ and $x_{dk}=F^{-1}\bigl(u_{dk};l(\pi_{dk}),\de_{dk}\bigr),\,d=0,1,\,k=1,\dots,K$.

The parameters  $\pi$, $(\pi_{11},\ldots,\pi_{1k},\ldots,\pi_{1K})$ $:=\bfpi_1$ and $(\pi_{01},\ldots,\pi_{0k},\ldots,\pi_{0K}):=\bfpi_0$ 
denote the meta-analytic parameters for the disease prevalence,  sensitivities and specificities, respectively, while the univariate parameters $\de$, $(\de_{11},$ $\ldots,\de_{1k},$
$\ldots,\de_{1K}):=\debf_1$ and $(\de_{01},\ldots,\de_{0k},$ $\ldots,\de_{0K}):=\debf_0$ 
denote the between-study variabilities  for the disease prevalence, sensitivities and specificities, respectively.
The parameter vector $\thbf$ 
of the random effects model  
is separated from the univariate parameters $\pi,\bfpi_1,\bfpi_0,\debf_1,\debf_0$ as   the random effects model is a 1-truncated C-vine copula model with dependence parameter vector $\thbf$  that does not involve the parameters of the within-studies model $\pi,\pibf_1, \pibf_0,\debf_1,\debf_0$.

We assume the same dependence structure for all studies. Therefore, studies reporting all test outcomes of $\T$ contribute to estimating the full dependence structure and studies with missing test outcomes directly contribute to estimating the marginal dependence structure. By assuming missing at random  (MAR) and the same dependence across all studies, which is equivalent to assuming all studies apply the multiple test comparison design, the 1-truncated C-vine copula mixed model  can combine studies reporting different sets of candidate tests and make inferences on the relative test performances.

\subsection{\label{relation2GLMM}Relationship with existing models}

When all the bivariate copulas $C_{dk,U}(\cdot;\th_{dk})$ are BVN   and the univariate distribution of the random effects is the  $N(\mu,\s)$ distribution, the between studies model is the  $(2K+1)$-variate normal model with  correlation  

$$\R=\begin{pmatrix}
1&\rho_{11} &\cdots&\rho_{1K}&\rho_{01} &\cdots&\rho_{0K}\\
\rho_{11}&1 &\cdots&\rho_{11,1K}&\rho_{11,01} &\cdots&\rho_{11,0K}\\
\vdots &\vdots& \ddots&\vdots&\vdots&\vdots&\vdots\\
\rho_{1K}&\rho_{11,1K}&\cdots&1&\rho_{1K,01} &\cdots&\rho_{1K,0K}\\
\rho_{01}&\rho_{11,01}  &\cdots&\rho_{1K,01}&1&\cdots&\rho_{01,0K}\\
\vdots &\ddots& \vdots&\vdots&\vdots&\ddots&\vdots\\
\rho_{0K}&\rho_{11,0K}  &\cdots&\rho_{1K,0K}&\rho_{01,0K}&\cdots&1
\end{pmatrix}$$

with 
\begin{equation}\label{BVN-rho}
\rho_{dk}=\th_{dk} \quad \mbox{and} \quad \rho_{d_1k_1,d_2k_2}=\theta_{d_1k_1}\theta_{d_2k_2}\quad \mbox{for} \quad d,d_1,d_2=0,1,\,k,k_1,k_2=1,\ldots,K.
\end{equation} 
This occurs because  the partial correlation  $\rho_{d_1k_1,d_2k_2|U}=\frac{\rho_{d_1k_1,d_2k_2}-\theta_{d_1k_1}\theta_{d_2k_2}}{\sqrt{1-\theta_{d_1k_1}^2}\sqrt{1-\theta_{d_2k_2}^2}}$ is zero due to the  assumption of conditional independence. As we assume conditional independence  a structured correlation matrix is exploited 
with $2K$ instead of $K(2K+1)$ correlation parameters.

The resulting random effects distribution for $(X,X_{11},\ldots,X_{1k},\ldots,X_{1K},X_{01},\ldots,X_{0k},\ldots,X_{0K})$ is the $2K+1$-variate normal distribution  with mean vector $\bmu=\bigl(l(\pi),l(\bfpi_1),l(\bfpi_0)\bigr)$ and variance-covariance matrix $\Sbf=\S \R \S$, where $\S$ is $(2K+1)\times (2K+1)$ diagonal matrix with diagonal elements  $(\s,\s_{11},\ldots,\s_{1K},\s_{01},\ldots,\s_{0K})$.
Hence,  the proposed model has as special case  the model in \cite{Ma-etal-2018-biostatistics}.

\subsection{Other choice of bivariate copulas}
Our general statistical model allows for selection of copulas and margins independently, i.e., there are no constraints in the choices of parametric copulas and margins. 

In line with our previous contributions in copula mixed models
\citep{Nikoloulopoulos2015b,Nikoloulopoulos2015c,Nikoloulopoulos-2016-SMMR,Nikoloulopoulos2018-AStA,nikoloulopoulos-2018-smmr,Nikoloulopoulos-2018-3dmeta-NE,Nikoloulopoulos-2018-4dmeta-NE,Nikoloulopoulos2020-factorREMADA,Nikoloulopoulos-2024,Nikoloulopoulos-2025-Biometrics}
we use 
 bivariate parametric copulas with different tail dependence behaviour, namely the bivariate normal (BVN) with intermediate tail dependence, Frank with tail independence, and Clayton with positive lower tail dependence. For the latter we also use its rotated versions to provide negative upper-lower tail dependence (Clayton rotated by 90$^\circ$), positive upper tail dependence (Clayton rotated by 180$^\circ$) and negative lower-upper tail dependence (Clayton rotated by 270$^\circ$).

\subsection{\label{computation}Maximum likelihood estimation and computational details}

Estimation of the model parameters $(\pi,\bfpi_1,\bfpi_0,\debf_1,\debf_0,\thbf)$   can be approached by the ML method, by maximizing the logarithm of the joint likelihood in (\ref{mixed-cop-likelihood}). 
The estimated parameters can be obtained by 
using a quasi-Newton \citep{nash90} method applied to the logarithm of the joint likelihood.  
This numerical  method requires only the objective
function, i.e.,  the logarithm of the joint likelihood, while the gradients
are computed numerically and the Hessian matrix of the second
order derivatives is updated in each iteration. The standard errors (SE) of the ML estimates can be also obtained via the gradients and the Hessian computed numerically during the maximization process.

For 1-truncated C-vine copula mixed models of the form with joint likelihood as in (\ref{mixed-cop-likelihood}), numerical evaluation of the joint pmf can be achieved with the following steps:

\begin{enumerate}
\itemsep=0pt
\item Calculate Gauss-Legendre \citep{Stroud&Secrest1966}
quadrature points $\{u_q: q=1,\ldots,N_q\}$ 
and weights $\{w_q: q=1,\ldots,N_q\}$ in terms of standard uniform.
\item Convert from independent uniform random variables $\{u_{q}: q=1,\ldots,N_q\}$, $\{u_{q_{11}}: q_{11}=1,\ldots,N_q\}$,  \ldots, $\{u_{q_{1K}}: q_{1K}=1,\ldots,N_q\}$, $\{u_{q_{01}}: q_{01}=1,\ldots,N_q\},\ldots,\{u_{q_{0K}}: q_{0K}=1,\ldots,N_q\}$,  to   dependent uniform random variables that have an 1-truncated C-vine distribution $C(\cdot;\thbf)$:  
\begin{algorithmic}[1]
\State   $v_{q}=u_{q}$
\For {$k=1,\ldots,K$} 
\State $v_{q_{1k}}=C_{1k}^{-1}
(u_{q_{1k}}|v_{q};\th_{1k})$
\State $v_{q_{0k}}=C_{0k}^{-1}
(u_{q_{0k}}|v_{q};\th_{0k})$
 \EndFor
\end{algorithmic}
where   $C_{dk}^{-1}(u_{q_{dk}}|v_{q};\th_{dk})$ is the  inverse  conditional bivariate copula cdf. The method is  based on the simulation algorithm of an 1-truncated C-vine copula \citep[Chapter 6]{joe2014}, where as input, instead of independent uniform variates, it uses the independent quadrature points.

\item Numerically evaluate the joint pmf
\begin{equation*}
\int_{[0,1]^{2K+1}}
\prod_{j=1}^{J_i}g\Bigl(\y_{ij},y_{ij0},\delta_{ij0}\mid x,\x_{1},\x_{0}\Bigr)c(u,u_{11},\ldots, u_{1K},u_{01},\ldots,u_{0K};\thbf)dud\u_{1}d\u_{0},\end{equation*}
where $\u_d=(u_{d1},\ldots,u_{dk}),\,d=0,1$,
in a multiple sum:
\begin{align*}
&\sum_{q=1}^{N_q}\sum_{q_{11}=1}^{N_q}\cdots\sum_{q_{1K}=1}^{N_q}
\sum_{q_{01}=1}^{N_q}\cdots\sum_{q_{0K}=1}^{N_q}w_{q}w_{q_{11}}\cdots w_{q_{1K}}w_{q_{01}}\cdots w_{q_{0K}}
g\Bigl(\y_{ij},y_{ij0},\delta_{ij0}\mid x,\x_{1},\x_{0}\Bigr),
\end{align*}
where  $x=F^{-1}\bigl(v_q;l(\pi),\de\bigr)$ and $x_{dk}=F^{-1}\bigl(v_{q_{dk}};l(\pi_{dk}),\de_{dk}\bigr),\,d=0,1,\,k=1,\dots,K$.

\end{enumerate}

With Gauss-Legendre quadrature, the same nodes and weights
are used for different functions;
this helps in yielding smooth numerical derivatives for numerical optimization via quasi-Newton.

\section{\label{miss-section}Small-sample efficiency -- Misspecification  of the parametric margin or bivariate pair-copula}
In this section, we study the small-sample efficiency and robustness of the ML estimation of the 1-truncated   C-vine copula mixed model. We gauge the small-sample efficiency of the ML method in Section \ref{computation} and investigate the misspecification of either  the parametric margin or bivariate copula of the random effects distribution.

We randomly generate 
1,000 meta-analysis data sets 
with $N =40$ studies in each data set. As in \cite{Ma-etal-2018-biostatistics}, we assume $K=2$, i.e., the whole test set contains two candidate tests ($T_1$ and $T_2$) and a gold standard ($T_0$). 
The missing indicators for all studies are prespecified such that the first $N_1=10$ studies are missing $T_2$,  the next $N_2=10$ studies are missing $T_1$, the next $N_3=10$ studies are missing $T_0$ (i.e., therein is an absence of the gold standard), and the last $N_4=10$  studies  do not have a missing test outcome.

To generate the data we have combined the simulation algorithms in \cite{Nikoloulopoulos2015b,nikoloulopoulos-2018-smmr,Nikoloulopoulos-2025-Biometrics}:

\begin{enumerate}
\item For $i=1,\ldots,N$, simulate $(u,u_{11},u_{12},u_{01},u_{02})$ from a 5-variate 1-truncated C-vine distribution $C_5(\cdot;$ $\tau_{11},\tau_{12},\tau_{01},\tau_{02})$ via the algorithm  in \citet[Chapter 6]{joe2014};  $\tau$'s are converted 
to $\theta$'s via the relations  in \citet[Chapter 4]{joe2014}. 
\item Convert to  normal or beta realizations   via $x=l^{-1}\bigl(F^{-1}(u;l(\pi),\de\bigr)$ and  $x_{dk}=l^{-1}\bigl(F^{-1}(u_{dk};$ $l(\pi_{dk}),\de_{dk}\bigr),\,d=0,1,\,k=1,2$.

\item Simulate the study size   $n$  from a shifted gamma distribution   with shape of 1.2, rate of 0.01 and shift of 30 to obtain heterogeneous study sizes \citep{paul-etal-2010}, 
and round off  $n$ to the nearest integers.

\item Set $i_1=[1,\ldots,N_1]$:

\begin{enumerate}

\item Draw the numbers of diseased $n_{1}$ from a $B\bigl(n[i_1],x[i_1]\bigr)$ distribution.
\item Set  the numbers of non-diseased as $n_0=n[i_1]-n_1$.

\item Generate   the numbers of diseased subjects $\sum_j y_{i_1j1}y_{i_1j0}$ with a positive result for $T_1$ 
from a 
$B\bigl(n_1,x_{11}[i_1]\bigr)$.
\item Generate   the numbers of non-diseased subjects $\sum_j (1-y_{i_1j1})(1-y_{i_1j0})$ with a negative result for $T_1$ 
from a 
$B\bigl(n_0,x_{01}[i_1]\bigr)$.
 
\end{enumerate}

\item Set $i_2=[N_1+1,\ldots,N_1+N_2]$:

\begin{enumerate}

\item Draw the numbers of diseased $n_{1}$ from a $B\bigl(n[i_2],x[i_2]\bigr)$ distribution.
\item Set  the numbers of non-diseased as $n_0=n[i2]-n_1$.

\item Generate   the numbers of diseased subjects $\sum_j y_{i_2j2}y_{i_2j0}$ with a positive result for $T_2$ 
from a 
$B\bigl(n_1,x_{12}[i_2]\bigr)$.
\item Generate   the numbers of non-diseased subjects $\sum_j (1-y_{i_2j2})(1-y_{i_2j0})$ with a negative result for $T_2$ 
from a 
$B\bigl(n_0,x_{02}[i_2]\bigr)$.

\end{enumerate}

\item Set $i_3=[N_1+N_2+1,\ldots,N_1+N_2+N_3]$:

\begin{enumerate}

\item 
 Draw $(m_{i_311},m_{i_310},
m_{i_301},m_{i_300})$ from
$\mathcal{M}_4\bigl(n[i_3],p_{11},p_{10},p_{01},p_{00}\bigr)$,
where  $\mathcal{M}_T(n,$ $p_1, \dots,p_{d})$  is shorthand notation for the  multinomial distribution; $d$ is the number of cells, $n$ is the number of observations, and  $(p_1,\dots,p_d)$ with $p_1+\ldots+p_d = 1$ is the $d$-dimensional vector of success probabilities and 
\begin{small}
\begin{eqnarray*}\label{cell-prob}
p_{11}&=&l^{-1}(x[i_3])l^{-1}(x_{11}[i_3])l^{-1}(x_{12}[i_3])+\bigl\{1-l^{-1}(x[i_3])\bigl\}\bigl\{1-l^{-1}(x_{01}[i_3])\bigr\}\bigl\{1-l^{-1}(x_{02}[i_3])\bigr\}\nonumber\\
p_{10}&=&l^{-1}(x[i_3])l^{-1}(x_{11}[i_3])\bigl\{1-l^{-1}(x_{12}[i_3])\bigr\}+\bigl\{1-l^{-1}(x[i_3])\bigr\}\bigl\{1-l^{-1}(x_{01}[i_3])\bigr\}l^{-1}(x_{02}[i_3])\\
p_{01}&=&l^{-1}(x[i_3])\bigl\{1-l^{-1}x_{11}[i_3])\bigr\}l^{-1}(x_{12}[i_3])+\bigl\{1-l^{-1}
(x[i_3])\bigr\}l^{-1}(x_{01}[i_3])\bigl\{1-l^{-1}(x_{02}[i_3])\bigr\}.\nonumber
  \end{eqnarray*}
  \end{small}
\item Set   the numbers of subjects with a positive result for $T_1$ and positive result for $T_2$
as 
$$\sum_j y_{i_3j1}y_{i_3j2}=m_{i_311}.$$
\item Set   the number of  subjects with a negative result for $T_1$ and negative result for $T_2$ 
as 
$$\sum_j (1-y_{i_3j1})(1-y_{i_3j2})=m_{i_300}.$$ 
 
\item Set   the numbers of  subjects with a positive result for $T_1$ and negative result for $T_2$ 
as 
$$\sum_j y_{i_3j1}(1-y_{i_3j2})=m_{i_310}.$$ 

\item Set   the numbers of  subjects with a negative result for $T_1$ and positive result for $T_2$ 
as 
$$\sum_j (1-y_{i_3j1})y_{i_3j2}=m_{i_301}.$$

\end{enumerate}

\item Set $i_4=[N_1+N_2+N_3+1,\ldots,N]$:

\begin{enumerate}

\item Draw the numbers of diseased $n_{1}$ from a $B\bigl(n[i_4],x[i_4]\bigr)$ distribution.
\item Set  the numbers of non-diseased as $n_0=n[i_4]-n_1$. 

\item Generate   the numbers of diseased subjects $\sum_j y_{i_4j1}y_{i_4j0}$ with a positive result for $T_1$ 
from a 
$B\bigl(n_1,x_{11}[i_4]\bigr)$.
\item Generate   the numbers of non-diseased subjects $\sum_j (1-y_{i_4j1})(1-y_{i_4j0})$ with a negative result for $T_1$ 
from a 
$B\bigl(n_0,x_{01}[i_4]\bigr)$.

\item Generate   the numbers of diseased subjects $\sum_j y_{i_4j2}y_{i_4j0}$ with a positive result for $T_2$ 
from a 
$B\bigl(n_1,x_{12}[i_4]\bigr)$.
\item Generate   the number of non-diseased subjects $\sum_j (1-y_{i_4j2})(1-y_{i_4j0})$ with a negative result for $T_2$ 
from a 
$B\bigl(n_0,x_{02}[i_4]\bigr)$.  

\end{enumerate}

\end{enumerate}

We use the same true parameters as in \cite{Ma-etal-2018-biostatistics}. That is, 
the sensitivity $\pi_{11}$  (specificity $\pi_{01}$) of $T_1$ is 0.8 (0.9) and the sensitivity $\pi_{12}$  (specificity $\pi_{02}$) of $T_2$ is 0.6 (0.7). The overall true disease prevalence $\pi$ is 0.4. We assume the random effects have variabilities of $\{\s,\s_{11},\s_{01},\s_{12},\s_{02}\}=0.3$ (normal margins) or $\{\g,\g_{11},\g_{01},\g_{12},\g_{02}\}=0.05$  (beta margins).  The Kendall's tau associations between prevalence and sensitivities or specificities  are set to be $\{\tau_{11},\tau_{01},\tau_{12},\tau_{02}\}=0.3$; this numerical value of $\tau$ is equivalent to a BVN correlation parameter $\rho$ of 0.5 via the relation in \cite{HultLindskog02}.

Representative summaries of findings on the performance of the ML method in Section \ref{computation} are given in Table \ref{sim1} and  Table \ref{sim2}
for  1-truncated C-vine copula models with normal and beta margins, respectively.
The true (simulated) bivariate copulas are the Clayton   copulas rotated by $180^\circ$. 
We have estimated the 1-truncated C-vine copula  mixed model  with different bivariate  copulas and margins. 
Table \ref{sim1} and  Table \ref{sim2} contain the resultant biases, root mean square errors (RMSEs) and standard deviations (SDs), along with average standard errors (ASEs), scaled by 100, for the MLEs under different copula choices and margins. The standard errors of the MLEs are obtained via the gradients and the Hessian that were computed numerically during the maximization process.

\begin{landscape}
\begin{table}[!h]
\caption{\label{sim1}
Simulation results when 1,000 meta-analysis data sets are generated with $N = 40$ studies in each data set from the 1-truncated C-vine copula mixed model with Clayton  copulas  rotated by $180^\circ$ and  normal margins. 
 }
\begin{small}

\begin{center}

    \begin{tabular}{>{\rowmac}l>{\rowmac}l>{\rowmac}l>{\rowmac}c>{\rowmac}c>{\rowmac}c>{\rowmac}c>{\rowmac}c>{\rowmac}c>{\rowmac}c>{\rowmac}c>{\rowmac}c>{\rowmac}c>{\rowmac}c>{\rowmac}c>{\rowmac}c>{\rowmac}c<{\clearrow}}
     \toprule
    &    &  & $\pi=$ & $\pi_{11}=$ & $\pi_{12}=$ & $\pi_{01}=$ & $\pi_{02}=$ & $\s=$ & $\s_{11}=$ & $\s_{12}=$ & $\s_{01}=$ & $\s_{02}=$ & $\tau_{11}=$ & $\tau_{12}=$ & $\tau_{01}=$ & $\tau_{02}=$ \\ 
 
   &   Margin & Copula & $0.4$ & $0.8$ & $0.6$ & $0.9$ & $0.7$ & $0.3$ & $0.3$ & $0.3$ & $0.3$ & $0.3$ & $0.3$ & $0.3$ & $0.3$ & $0.3$ \\  \midrule
 
        Bias & normal & BVN & -0.03 & -0.01 & -0.06 & -0.14 & -0.14 & -3.39 & -4.50 & -3.40 & -5.51 & -4.43 & 15.10 & 12.06 & 5.00 & 6.20 \\ 

        ~ & ~ & Frank & 0.08 & 0.07 & 0.03 & -0.11 & -0.06 & -3.34 & -4.42 & -3.31 & -5.80 & -4.44 & 11.80 & 10.24 & 1.25 & 5.74 \\ 
        ~ & ~ & Clayton & -0.02 & -0.01 & -0.05 & -0.15 & -0.16 & -3.20 & -4.35 & -3.26 & -6.53 & -4.78 & 16.72 & 12.23 & 8.13 & 6.21 \\ 
 \setrow{\bfseries}         ~ & ~ & Cln180$^\circ$ & -0.02 & 0.02 & -0.03 & -0.11 & -0.11 & -3.25 & -5.09 & -3.85 & -6.03 & -4.24 & 13.42 & 10.05 & 7.04 & 7.04 \\ 
        ~ & beta & BVN & 0.14 & -0.36 & -0.23 & -0.40 & -0.42 & - & - & - & - & - & 15.69 & 13.06 & 5.59 & 6.60 \\ 

        ~ & ~ & Frank & 0.21 & -0.30 & -0.16 & -0.38 & -0.32 & - & - & - & - & - & 12.24 & 12.09 & 3.29 & 6.29 \\ 
        ~ & ~ & Clayton & 0.15 & -0.37 & -0.23 & -0.40 & -0.45 & - & - & - & - & - & 17.87 & 12.52 & 9.71 & 5.91 \\ 
        ~ & ~ & Cln180$^\circ$ & 0.16 & -0.32 & -0.19 & -0.37 & -0.39 & - & - & - & - & - & 16.25 & 12.29 & 9.35 & 8.86 \\ 
  \midrule       SD  & normal & BVN & 1.45 & 1.64 & 2.00 & 1.05 & 1.65 & 5.54 & 11.62 & 8.98 & 12.40 & 7.80 & 46.43 & 35.51 & 54.37 & 32.01 \\ 

         ~ & ~ & Frank & 1.48 & 1.65 & 2.01 & 1.05 & 1.66 & 5.56 & 11.54 & 8.96 & 12.14 & 7.72 & 41.36 & 32.21 & 49.67 & 30.23 \\ 
         ~ & ~ & Clayton & 1.46 & 1.66 & 2.01 & 1.05 & 1.66 & 5.60 & 12.43 & 9.49 & 13.56 & 8.34 & 38.04 & 35.39 & 36.97 & 32.98 \\ 
  \setrow{\bfseries}       ~ & ~ & Cln180$^\circ$ & 1.44 & 1.63 & 1.99 & 1.04 & 1.62 & 5.22 & 11.67 & 8.85 & 13.11 & 7.71 & 31.23 & 29.23 & 31.66 & 26.60 \\ 
      ~ & beta & BVN & 1.42 & 1.66 & 1.98 & 1.05 & 1.62 & 0.66 & 0.98 & 1.06 & 0.66 & 0.81 & 46.69 & 36.52 & 54.36 & 32.47 \\ 
 
       ~ & ~ & Frank & 1.42 & 1.65 & 1.98 & 1.05 & 1.62 & 0.66 & 0.98 & 1.04 & 0.66 & 0.80 & 42.43 & 35.07 & 50.24 & 32.53 \\ 
       ~ & ~ & Clayton & 1.42 & 1.67 & 1.99 & 1.05 & 1.63 & 0.68 & 1.04 & 1.13 & 0.65 & 0.83 & 38.19 & 35.70 & 37.35 & 32.48 \\ 
        ~ & ~ & Cln180$^\circ$ & 1.42 & 1.63 & 1.95 & 1.03 & 1.62 & 0.64 & 0.91 & 1.04 & 0.63 & 0.80 & 32.97 & 30.17 & 33.50 & 28.02 \\ \midrule
        ASE & normal & BVN & 1.27 & 1.57 & 1.93 & 0.97 & 1.53 & 4.64 & 11.25 & 8.40 & 12.68 & 7.36 & 59.52 & 35.77 & 103.57 & 30.32 \\ 
 
        ~ & ~ & Frank & 1.27 & 1.58 & 1.92 & 0.96 & 1.54 & 4.62 & 11.32 & 8.44 & 12.35 & 7.40 & 41.15 & 29.22 & 69.03 & 25.18 \\ 
        ~ & ~ & Clayton & 1.27 & 1.60 & 1.92 & 0.95 & 1.56 & 4.61 & 11.13 & 8.44 & 12.08 & 7.39 & 46.56 & 27.21 & 49.13 & 27.12 \\ 
\setrow{\bfseries}        ~ & ~ & Cln180$^\circ$ & 1.29 & 1.56 & 1.92 & 0.95 & 1.55 & 4.53 & 11.07 & 8.16 & 12.32 & 7.27 & 39.45 & 26.13 & 48.18 & 21.82 \\ 
        ~ & beta & BVN & 1.25 & 1.57 & 1.89 & 0.96 & 1.51 & 0.56 & 0.90 & 1.00 & 0.57 & 0.75 & 75.46 & 43.28 & 74.63 & 28.28 \\ 
 
        ~ & ~ & Frank & 1.25 & 1.58 & 1.88 & 0.96 & 1.51 & 0.56 & 0.90 & 1.00 & 0.57 & 0.76 & 42.93 & 27.22 & 60.30 & 28.93 \\ 
        ~ & ~ & Clayton & 1.25 & 1.56 & 1.88 & 0.94 & 1.54 & 0.56 & 0.92 & 1.01 & 0.57 & 0.76 & 36.31 & 25.85 & 54.93 & 23.04 \\ 
        ~ & ~ & Cln180$^\circ$ & 1.24 & 1.55 & 1.86 & 0.95 & 1.50 & 0.55 & 0.85 & 0.96 & 0.57 & 0.74 & 33.62 & 26.23 & 48.28 & 22.66 \\ 
   \midrule      RMSE & normal & BVN & 1.45 & 1.64 & 2.00 & 1.05 & 1.65 & 6.50 & 12.46 & 9.60 & 13.57 & 8.97 & 48.83 & 37.50 & 54.60 & 32.60 \\ 

        ~ & ~ & Frank & 1.48 & 1.65 & 2.01 & 1.05 & 1.67 & 6.48 & 12.36 & 9.55 & 13.45 & 8.90 & 43.01 & 33.80 & 49.69 & 30.77 \\ 
         ~ & ~ & Clayton & 1.46 & 1.66 & 2.01 & 1.06 & 1.66 & 6.45 & 13.17 & 10.04 & 15.05 & 9.61 & 41.55 & 37.44 & 37.85 & 33.56 \\ 
   \setrow{\bfseries}     ~ & ~ & Cln180$^\circ$ & 1.44 & 1.63 & 1.99 & 1.04 & 1.63 & 6.15 & 12.73 & 9.65 & 14.43 & 8.80 & 33.99 & 30.91 & 32.43 & 27.52 \\ 
       ~ & beta & BVN & 1.43 & 1.69 & 1.99 & 1.13 & 1.68 & - & - & - & - & - & 49.26 & 38.78 & 54.65 & 33.13 \\ 

        ~ & ~ & Frank & 1.44 & 1.67 & 1.98 & 1.11 & 1.65 & - & - & - & - & - & 44.16 & 37.10 & 50.35 & 33.14 \\ 
     ~ & ~ & Clayton & 1.43 & 1.71 & 2.01 & 1.13 & 1.69 & - & - & - & - & - & 42.16 & 37.83 & 38.60 & 33.01 \\ 
       ~ & ~ & Cln180$^\circ$ & 1.43 & 1.66 & 1.96 & 1.10 & 1.67 & - & - & - & - & - & 36.76 & 32.57 & 34.78 & 29.39 \\ \bottomrule
    \end{tabular}
    \end{center}
\end{small}

  \begin{flushleft}
\begin{footnotesize}
The biases,  root mean square errors (RMSEs) and standard deviations (SDs), along with average standard errors (ASEs) for the MLEs  are boldfaced under the true model   and  scaled by 100 under the various copula choices and margins; 
Cln$180^\circ$: Clayton  copula rotated by $180^\circ$; 
$N_q=15$ quadrature points have been used. 
\end{footnotesize}  
\end{flushleft}    
\end{table}
\end{landscape}

\begin{landscape}
\begin{table}[!h]
\caption{\label{sim2}
Simulation results when 1,000 meta-analysis data sets are generated with $N = 40$ studies in each data set from the 1-truncated C-vine copula mixed model with Clayton  copulas  rotated by $180^\circ$ and  beta margins. 
 }
\begin{small}

\begin{center}

    \begin{tabular}{>{\rowmac}l>{\rowmac}l>{\rowmac}l>{\rowmac}c>{\rowmac}c>{\rowmac}c>{\rowmac}c>{\rowmac}c>{\rowmac}c>{\rowmac}c>{\rowmac}c>{\rowmac}c>{\rowmac}c>{\rowmac}c>{\rowmac}c>{\rowmac}c>{\rowmac}c<{\clearrow}}
    \toprule
   &      & & $\pi=$ & $\pi_{11}=$ & $\pi_{12}=$ & $\pi_{01}=$ & $\pi_{02}=$ & $\g=$ & $\g_{11}=$ & $\g_{12}=$ & $\g_{01}=$ & $\g_{02}=$ & $\tau_{11}=$ & $\tau_{12}=$ & $\tau_{01}=$ & $\tau_{02}=$ \\ 

   &     Margin & Copula & $0.4$ & $0.8$ & $0.6$ & $0.9$ & $0.7$ & $0.05$ & $0.05$ & $0.05$ & $0.05$ & $0.05$ & $0.3$ & $0.3$ & $0.3$ & $0.3$ \\ \midrule
        Bias & normal & BVN & -0.57 & 1.20 & 0.39 & 1.40 & 0.62 & - & - & - & - & - & 8.69 & 7.96 & 0.15 & 2.57 \\ 

        ~ & ~ & Frank & -0.11 & 1.46 & 0.75 & 1.49 & 0.85 & - & - & - & - & - & 7.96 & 7.76 & 0.16 & 3.08 \\ 
        ~ & ~ & Clayton & -0.45 & 1.27 & 0.48 & 1.35 & 0.61 & - & - & - & - & - & 6.62 & 5.71 & -2.63 & -1.70 \\ 
        ~ & ~ & Cln180$^\circ$ & -0.52 & 1.30 & 0.44 & 1.48 & 0.64 & - & - & - & - & - & 7.01 & 5.83 & 1.76 & 2.99 \\ 
        ~ & beta & BVN & -0.11 & -0.10 & -0.04 & -0.11 & -0.21 & -0.76 & -0.65 & -0.65 & -0.86 & -0.82 & 8.98 & 7.89 & -0.55 & 2.15 \\ 

        ~ & ~ & Frank & 0.27 & 0.18 & 0.27 & 0.04 & -0.03 & -0.72 & -0.68 & -0.62 & -0.97 & -0.85 & 8.67 & 7.68 & -0.38 & 2.88 \\ 
        ~ & ~ & Clayton & 0.02 & -0.11 & 0.02 & -0.16 & -0.24 & -0.59 & -0.27 & -0.35 & -0.84 & -0.79 & 6.97 & 5.34 & -3.64 & -2.26 \\ 
 \setrow{\bfseries}       ~ & ~ & Cln180$^\circ$ & -0.07 & 0.05 & 0.05 & -0.03 & -0.19 & -0.86 & -0.94 & -0.81 & -0.94 & -0.83 & 7.53 & 6.35 & 1.53 & 2.68 \\ \midrule
         SD  & normal & BVN & 2.15 & 2.44 & 2.72 & 1.67 & 2.32 & 7.63 & 16.33 & 11.44 & 19.48 & 10.09 & 25.96 & 24.34 & 24.86 & 20.37 \\ 

         ~ & ~ & Frank & 2.25 & 2.46 & 2.78 & 1.70 & 2.39 & 7.72 & 16.60 & 11.55 & 19.24 & 10.08 & 25.93 & 24.25 & 24.52 & 20.50 \\ 
        ~ & ~ & Clayton & 2.19 & 2.52 & 2.78 & 1.71 & 2.34 & 8.14 & 18.04 & 12.53 & 20.49 & 10.64 & 29.05 & 27.91 & 23.29 & 21.63 \\ 
        ~ & ~ & Cln180$^\circ$ & 2.17 & 2.42 & 2.70 & 1.67 & 2.34 & 7.26 & 15.42 & 11.15 & 19.15 & 9.90 & 22.91 & 21.69 & 21.87 & 19.07 \\ 
        ~ & beta & BVN & 2.07 & 2.44 & 2.63 & 1.73 & 2.23 & 1.35 & 2.32 & 2.01 & 2.11 & 1.62 & 26.18 & 24.54 & 24.64 & 20.07 \\ 

         ~ & ~ & Frank & 2.15 & 2.41 & 2.66 & 1.72 & 2.29 & 1.38 & 2.27 & 2.03 & 2.04 & 1.62 & 26.05 & 24.55 & 24.58 & 20.52 \\ 
        ~ & ~ & Clayton & 2.10 & 2.50 & 2.68 & 1.73 & 2.25 & 1.47 & 2.64 & 2.25 & 2.21 & 1.73 & 29.39 & 27.87 & 22.97 & 21.37 \\ 
  \setrow{\bfseries}      ~ & ~ & Cln180$^\circ$ & 2.07 & 2.38 & 2.60 & 1.72 & 2.25 & 1.27 & 2.15 & 1.92 & 2.02 & 1.59 & 23.34 & 22.63 & 21.64 & 18.87 \\ \midrule
        ASE & normal & BVN & 1.83 & 2.28 & 2.60 & 1.50 & 2.20 & 5.90 & 14.31 & 9.96 & 17.39 & 9.38 & 22.11 & 19.89 & 23.04 & 17.04 \\ 
 
        ~ & ~ & Frank & 1.82 & 2.25 & 2.59 & 1.49 & 2.20 & 5.99 & 14.47 & 10.05 & 17.32 & 9.38 & 21.24 & 19.65 & 21.64 & 17.40 \\ 
        ~ & ~ & Clayton & 1.86 & 2.40 & 2.66 & 1.50 & 2.21 & 6.11 & 15.06 & 10.29 & 17.17 & 9.36 & 19.34 & 18.98 & 16.52 & 15.78 \\ 
        ~ & ~ & Cln180 & 1.79 & 2.23 & 2.55 & 1.47 & 2.18 & 5.60 & 13.79 & 9.53 & 16.89 & 9.15 & 19.32 & 17.75 & 18.40 & 16.09 \\ 
        ~ & beta & BVN & 1.75 & 2.27 & 2.49 & 1.57 & 2.11 & 1.05 & 2.06 & 1.76 & 1.87 & 1.53 & 22.35 & 19.86 & 20.47 & 16.99 \\ 
 
        ~ & ~ & Frank & 1.75 & 2.26 & 2.48 & 1.56 & 2.10 & 1.07 & 2.09 & 1.78 & 1.86 & 1.53 & 21.58 & 19.72 & 20.78 & 17.40 \\ 
        ~ & ~ & Clayton & 1.76 & 2.33 & 2.53 & 1.56 & 2.10 & 1.09 & 2.22 & 1.87 & 1.86 & 1.54 & 20.03 & 18.86 & 16.81 & 15.65 \\ 
  \setrow{\bfseries}      ~ & ~ & Cln180$^\circ$ & 1.71 & 2.23 & 2.43 & 1.56 & 2.09 & 0.99 & 1.92 & 1.65 & 1.82 & 1.50 & 19.39 & 17.96 & 18.15 & 16.08 \\ \midrule
        RMSE & normal & BVN & 2.23 & 2.72 & 2.75 & 2.18 & 2.40 & - & - & - & - & - & 27.38 & 25.61 & 24.86 & 20.53 \\ 

          ~ & ~ & Frank & 2.25 & 2.86 & 2.88 & 2.26 & 2.54 & - & - & - & - & - & 27.12 & 25.46 & 24.52 & 20.73 \\ 
       ~ & ~ & Clayton & 2.24 & 2.82 & 2.82 & 2.18 & 2.42 & - & - & - & - & - & 29.80 & 28.49 & 23.44 & 21.69 \\ 
      ~ & ~ & Cln180 & 2.23 & 2.75 & 2.73 & 2.23 & 2.42 & - & - & - & - & - & 23.96 & 22.46 & 21.94 & 19.30 \\ 
       ~ & beta & BVN & 2.08 & 2.44 & 2.63 & 1.74 & 2.24 & 1.55 & 2.41 & 2.11 & 2.28 & 1.81 & 27.67 & 25.78 & 24.65 & 20.18 \\ 

         ~ & ~ & Frank & 2.17 & 2.41 & 2.67 & 1.72 & 2.29 & 1.55 & 2.37 & 2.12 & 2.26 & 1.83 & 27.46 & 25.72 & 24.58 & 20.72 \\ 
         ~ & ~ & Clayton & 2.10 & 2.50 & 2.68 & 1.74 & 2.27 & 1.58 & 2.65 & 2.27 & 2.37 & 1.90 & 30.20 & 28.38 & 23.25 & 21.48 \\ 
   \setrow{\bfseries}      ~ & ~ & Cln180$^\circ$ & 2.07 & 2.38 & 2.60 & 1.72 & 2.26 & 1.53 & 2.35 & 2.08 & 2.23 & 1.80 & 24.53 & 23.50 & 21.69 & 19.06 \\ \bottomrule
    \end{tabular}

\end{center}
\end{small}
  \begin{flushleft}
\begin{footnotesize}
The biases,  root mean square errors (RMSEs) and standard deviations (SDs), along with average standard errors (ASEs) for the MLEs  are boldfaced under the true model   and  scaled by 100 under the various copula choices and margins; 
Cln$180^\circ$: Clayton  copula rotated by $180^\circ$; 
$N_q=15$ quadrature points have been used. 
\end{footnotesize}  
\end{flushleft}    
\end{table}
\end{landscape}

From the summaries above, ML   with  the true 1-truncated C-vine copula mixed  model is highly efficient according to the simulated biases,  SDs and RMSEs for the parameters of main interest $\pi,\bfpi_1,\bfpi_0$. However, the MLEs  of $\pi,\bfpi_1,\bfpi_0$ are not robust to margin misspecification, e.g.,  in Table \ref{sim1} (Table \ref{sim2}) where the true univariate margins are normal (beta) the scaled biases for the MLEs of $\pi_{01}$ for the various 1-truncated C-vine copula mixed models with beta (normal) margins range from $-0.40$ ($1.35$) to $-0.37$ (1.49).

We also show
that the  ML  estimates of $\tau$'s   are robust to  margin misspecification, as the copula remains invariant under any series of strictly increasing transformations of the components of the random vector. 
We  observe that  larger scaled 
biases, SDs, ASEs and RMSEs  are introduced for the Kendall's $\tau$ and variability parameters   $\s,\sbf_1,\sbf_0$ (normal margins) or $\g,\gbf_1,\gbf_0$ (beta margins). This is because five variability and four Kendall's $\tau$ parameters have to be estimated in addition to  the five probability parameters that are of  main interest. 
Nevertheless, this does not have implications for the parameters of  main interest $\pi, \pibf_1,\pibf_0$,  i.e.,  the meta-analytic parameters for the disease prevalence,  sensitivities and specificities, respectively.

The simulation results indicate that  the effect of misspecifying the marginal choice can be seen as substantial for the univariate parameters of  main interest $\pi, \pibf_1,\pibf_0$,  i.e., the meta-analytic parameters   for the disease prevalence,  sensitivities and specificities, respectively. Hence, the model in \cite{Ma-etal-2018-biostatistics} can lead to biased meta-analytic estimates of interest $\pi, \pibf_1,\pibf_0$ as it is restricted to a normal margin specification.

Last but not least, the effect of misspecifying the copula choice can be seen as minimal for both the univariate parameters and Kendall's tau, which  is a strictly increasing function of the copula parameter for any pair-copula,  as  (a) the meta-analytic parameters are a univariate inference, and hence, it is the univariate marginal distribution that matters and not the type of the pair-copula, and (b) Kendall's tau only accounts for the dependence dominated by the middle of the data, and it is expected to be similar amongst different families of bivariate copulas. However, the tail dependence varies, and is a property to consider when choosing amongst different families of bivariate copulas. Any inference that depends on the joint distribution  will essentially show the effects of different model (random effect distribution) assumptions such as the pair-copula choice.

\section{\label{app-section}Application}
Deep vein thrombosis (DVT) is characterized by the formation of blood clots within one or more deep veins of the body. If untreated, DVT can progress to pulmonary embolism, a potentially fatal complication \citep{Venta1987}. Contrast venography remains the gold-standard diagnostic modality for DVT; however, its invasive nature and the associated risk of allergic reactions limit its clinical applicability. Consequently, ultrasonography is widely employed as a noninvasive diagnostic alternative with high accuracy. Additionally, measurement of plasma D-dimer levels, representing fibrin degradation products indicative of intravascular coagulation, can also serve as a convenient screening tool for DVT diagnosis.

\cite{Kang-etal-2013} conducted a meta-analysis of $N=12$ studies evaluating the diagnostic performance of these modalities. Of the included studies, $N_1=4$ compared D-dimer testing with venography, $N_2=3$ compared ultrasonography with venography, and $N_3=5$ compared D-dimer testing with ultrasonography. None of the $N=12$ studies directly compared all three diagnostic methods concurrently, i.e., $N_4=0$. 
We demonstrate the modelling process of the proposed approach 
 by insightfully re-analyzing these data. These data have previously been analysed by \cite{Ma-etal-2018-biostatistics} and  \cite{Lian-eta-al-2019-jasa}.
Note in passing that the  model in  \cite{Ma-etal-2018-biostatistics} is a special case of our model when all the bivariate copulas are BVN  and the univariate distribution of the random effects is 
the $N(\mu,\s^2)$ distribution as shown in Section \ref{relation2GLMM}.

\begin{table}[!ht]
\caption{\label{app-table}Maximized log-likelihoods, estimates and standard errors (SE)  of the  1-truncated C-vine copula mixed models with normal  and beta margins for the the meta-analysis of $N=12$ studies for  the diagnostic modality for deep vein thrombosis;   $N_1=4$ compared D-dimer ($T_1$) testing with venography ($T_0$), $N_2=3$ compared ultrasonography ($T_2$) with venography ($T_0$), and $N_3=5$ compared D-dimer ($T_1$) testing with ultrasonography ($T_2$).}  
    \centering
    \begin{tabular}{ccccccccccccc}
    \toprule
 &&        \multicolumn{2}{c}{BVN}& ~ &  \multicolumn{2}{c}{Frank}& ~ &  \multicolumn{2}{c}{Clayton} & ~ &  \multicolumn{2}{c}{Cln180$^\circ$}  \\ 
        ~ & ~ & Est. & SE & ~ & Est. & SE & ~ & Est. & SE & ~ & Est. & SE \\ 
         \multicolumn{13}{l}{Normal margins}\\ \cmidrule{3-4}\cmidrule{6-7}\cmidrule{9-10}\cmidrule{12-13}
        $\pi$ & ~ & 0.438 & 0.025 & ~ & 0.445 & 0.026 & ~ & 0.435 & 0.028 & ~ & 0.439 & 0.026 \\ 
        $\pi_{11}$ & ~ & 0.835 & 0.052 & ~ & 0.842 & 0.052 & ~ & 0.833 & 0.055 & ~ & 0.836 & 0.052 \\ 
        $\pi_{12}$ & ~ & 0.955 & 0.041 & ~ & 0.965 & 0.047 & ~ & 0.953 & 0.045 & ~ & 0.957 & 0.049 \\ 
        $\pi_{01}$ & ~ & 0.879 & 0.054 & ~ & 0.881 & 0.054 & ~ & 0.891 & 0.055 & ~ & 0.873 & 0.053 \\ 
        $\pi_{02}$ & ~ & 0.842 & 0.098 & ~ & 0.863 & 0.104 & ~ & 0.823 & 0.191 & ~ & 0.825 & 0.124 \\ 
        $\s$ & ~ & 0.254 & 0.105 & ~ & 0.247 & 0.100 & ~ & 0.254 & 0.136 & ~ & 0.244 & 0.096 \\ 
        $\s_{11}$ & ~ & 0.918 & 0.390 & ~ & 0.909 & 0.389 & ~ & 0.903 & 0.421 & ~ & 0.913 & 0.385 \\ 
        $\s_{12}$ & ~ & 1.351 & 0.810 & ~ & 1.410 & 0.919 & ~ & 1.441 & 0.978 & ~ & 1.319 & 0.908 \\ 
        $\s_{01}$ & ~ & 1.076 & 0.501 & ~ & 1.022 & 0.448 & ~ & 1.053 & 0.491 & ~ & 0.977 & 0.395 \\ 
        $\s_{02}$ & ~ & 2.140 & 0.957 & ~ & 2.010 & 0.876 & ~ & 2.562 & 1.168 & ~ & 1.887 & 0.866 \\ 
        $\tau_{11}$ & ~ & 0.950 & - & ~ & 0.950 & - & ~ & 0.950 & - & ~ & 0.950 & - \\ 
        $\tau_{12}$ & ~ & 0.365 & 0.347 & ~ & 0.427 & 0.464 & ~ & 0.299 & 0.410 & ~ & 0.474 & 0.458 \\ 
        $\tau_{01}$ & ~ & 0.189 & 0.334 & ~ & 0.115 & 0.338 & ~ & 0.000 & 0.787 & ~ & 0.207 & 0.290 \\ 
        $\tau_{02}$ & ~ & 0.533 & 0.081 & ~ & 0.580 & 0.262 & ~ & 0.558 & 0.131 & ~ & 0.635 & 0.174 \\ \midrule
        $-\ell$ & ~ &  \multicolumn{2}{c}{1063.23}& ~ &  \multicolumn{2}{c}{1062.92}& ~ &  \multicolumn{2}{c}{1064.79}& ~ &  \multicolumn{2}{c}{1062.09}\\ \midrule
       Beta margins & ~ & ~ & ~ & ~ & ~ & ~ & ~ & ~ & ~ & ~ & ~ & ~ \\ 
     $\pi$ & ~ & 0.440 & 0.026 & ~ & 0.446 & 0.026 & ~ & 0.440 & 0.028 & ~ & 0.439 & 0.025 \\ 
     $\pi_{11}$ & ~ & 0.804 & 0.051 & ~ & 0.810 & 0.050 & ~ & 0.807 & 0.050 & ~ & 0.802 & 0.052 \\ 
        $\pi_{12}$ & ~ & 0.919 & 0.043 & ~ & 0.930 & 0.046 & ~ & 0.923 & 0.044 & ~ & 0.924 & 0.046 \\ 
        $\pi_{01}$ & ~ & 0.839 & 0.052 & ~ & 0.841 & 0.050 & ~ & 0.844 & 0.052 & ~ & 0.833 & 0.049 \\ 
        $\pi_{02}$ & ~ & 0.742 & 0.083 & ~ & 0.763 & 0.082 & ~ & 0.780 & 0.111 & ~ & 0.732 & 0.086 \\ 
        $\g$ & ~ & 0.015 & 0.012 & ~ & 0.015 & 0.012 & ~ & 0.015 & 0.013 & ~ & 0.014 & 0.011 \\ 
       $\g_{11}$ & ~ & 0.103 & 0.071 & ~ & 0.102 & 0.071 & ~ & 0.102 & 0.074 & ~ & 0.100 & 0.069 \\ 
        $\g_{12}$ & ~ & 0.112 & 0.104 & ~ & 0.122 & 0.115 & ~ & 0.115 & 0.106 & ~ & 0.123 & 0.119 \\ 
         $\g_{01}$ & ~ & 0.129 & 0.089 & ~ & 0.119 & 0.077 & ~ & 0.127 & 0.086 & ~ & 0.106 & 0.065 \\ 
       $\g_{02}$ & ~ & 0.363 & 0.153 & ~ & 0.345 & 0.160 & ~ & 0.361 & 0.116 & ~ & 0.333 & 0.144 \\ 
     $\tau_{11}$ & ~ & 0.950 & - & ~ & 0.950 & - & ~ & 0.950 & - & ~ & 0.950 & - \\ 
        $\tau_{12}$ & ~ & 0.348 & 0.353 & ~ & 0.425 & 0.423 & ~ & 0.241 & 0.379 & ~ & 0.497 & 0.476 \\ 
         $\tau_{01}$ & ~ & 0.173 & 0.327 & ~ & 0.109 & 0.320 & ~ & 0.000 & 0.516 & ~ & 0.162 & 0.286 \\ 
         $\tau_{02}$ & ~ & 0.478 & 0.125 & ~ & 0.566 & 0.199 & ~ & 0.304 & 0.070 & ~ & 0.614 & 0.156 \\ \midrule
        $-\ell$ & ~ &  \multicolumn{2}{c}{1062.76}& ~ &  \multicolumn{2}{c}{1062.37}& ~ &  \multicolumn{2}{c}{1064.22} & ~ &  \multicolumn{2}{c}{1061.60} \\ \bottomrule
    \end{tabular}
       \begin{flushleft}
\begin{footnotesize}
The resulting model with normal margins and BVN copulas is   the model in \cite{Ma-etal-2018-biostatistics}; Cln$180^\circ$: Clayton  copula rotated by $180^\circ$; 
$N_q=15$ quadrature points have been used.
\end{footnotesize} 
\end{flushleft} 
\end{table}

We fit the 1-truncated C-vine copula mixed model for all different  pair copulas and univariate marginal distributions. 
In our  general statistical model, 
there are no constraints in the choice of the parametric marginal  or pair-copula  distributions.
This is one of the limitations of the  the model in \cite{Ma-etal-2018-biostatistics} where all the pair copulas are BVN and marginal distributions are normal.  However, for ease of interpretation, we do not mix pair-copulas or margins.
To make it easier
to compare strengths of dependence amongst different copulas, we convert from the BVN, Frank  and   (rotated)  Clayton
$\theta$'s to  
to $\tau$'s via the relations  in \citet[Chapter 4]{joe2014}.
 In cases when fitting the 1-truncated C-vine copula mixed model, the resultant estimate of one of the  Kendall's $\tau$ parameters was close to  the right ($0.95$) or left boundary ($-0.95$) of its parameter space, we set the corresponding bivariate copula to comonotonic (Fr\'echet upper bound) or countermonotonic  (Fr\'echet lower bound) copula, respectively.

To identify the best-fitting model, we do not rely on formal goodness-of-fit tests. Instead, we use the log-likelihood evaluated at the maximum likelihood estimate as a heuristic measure for comparing model fit. Traditional goodness-of-fit procedures are based on a global distance between the empirical and model-implied distributions and may therefore lack sensitivity to tail behavior. Moreover, small 
$p$-values from these tests are not diagnostically informative, as they do not suggest how the parametric model might be improved \citep[page 254]{joe2014}. In the context of vine copulas, \cite{Dissmann-etal-2013-csda} show that likelihood-based pair-copula selection outperforms approaches based on bivariate goodness-of-fit tests. Higher likelihood values indicate models that more accurately capture both the overall dependence structure and the strength of tail dependence in the data.

The maximized log-likelihoods, estimates and standard errors
from fitting  the 1-truncated C-vine copula mixed models   are given  in   Table \ref{app-table}. 
The log-likelihoods show that a  1-truncated C-vine copula mixed model with Clayton rotated by $180^\circ$   bivariate copulas and beta margins provides the best fit.  
It is revealed that a 1-truncated C-vine copula mixed model with the disease prevalence, sensitivities and specificities  on the original scale provides better  fit than the  model in \cite{Ma-etal-2018-biostatistics}, which models them on a transformed scale. The improvement over the model in \cite{Ma-etal-2018-biostatistics} is small in terms of the likelihood principle, but for a sample size such as $N=12$, $-1061.60-(-1063.23)=1.63$ units log-likelihood difference is sufficient.

The fact that the best-fitting bivariate copulas are Clayton rotated by 180$^\circ$ reveals that there exists upper tail dependence amongst the disease prevalence, sensitivities and specificities. 
It is  also apparent that the estimates of the meta-analytic parameters  of sensitivity and specificity for each test  from the model in \cite{Ma-etal-2018-biostatistics} differentiate from the ones from the selected model with beta margins.  The meta-analytic parameters of sensitivity and specificity for each test 
are upward biased  when  normal margins are assumed. 
This is consistent  with the  simulation results and conclusions  in Section \ref{miss-section} when the univariate random effects are misspecified.
 Our general model can allow both normal and beta margins, i.e., it is not restricted to normal margins as the model in \cite{Ma-etal-2018-biostatistics}.

\section{\label{discussion}Discussion}

We have proposed a 1-truncated C-vine copula mixed model for network  meta-analysis  of  multiple diagnostic tests.  Our  model 
generalizes 
the model in \cite{Ma-etal-2018-biostatistics}
that can lead to biased estimates of the meta-analytic parameters of interest. 
It essentially provides an improvement  as  the random effects distribution is expressed via a vine copula that allows for flexible dependence modelling, different from assuming simple linear correlation structures and  normality. This strength of multivariate meta-analysis  approaches that use copulas has been  pointed out
\citep{Jacson&White-2018-BiomJ,jackson&white&riley-2020} and it has also  been exploited 
 in network  meta-analysis of interventions
 \citep{Phillippo-etal-2020-jrssa}. 
Vine copulas, by choosing bivariate  copulas appropriately,  can have a flexible range of lower/upper tail dependence \citep{joeetal10}. The 1-truncated C-vine copula mixed model allows for selection of parametric bivariate copulas and univariate margins independently among a variety of parametric families. Hence, the latent prevalence, sensitivities and specificities can be modelled on the original proportions scale and can be tail dependent.

We propose an efficient  ML estimation technique based on dependent  Gauss-Legendre quadrature points that have a 1-truncated C-vine copula distribution.  
 We use the notion of a truncated at level 1 C-vine copula that leads to a substantial reduction  of the dependence parameters.  This is extremely useful for estimation purposes  given the typical small sample sizes in meta-analysis of diagnostic test accuracy studies.  
\cite{Ma-etal-2018-biostatistics} perform estimation using MCMC methods in the Bayesian framework and acknowledge that optimizing the likelihood for joint meta-analysis is non-trivial, because it involves calculating complicated integrals numerically.
Our numerical method  that is based on dependent  Gauss-Legendre quadrature points, that have a 1-truncated C-vine copula distribution, successively computes the $(2K+1)$-dimensional integrals in  multiple sums over the dependent quadrature points and weights.

In an era of evidence-based medicine, decision makers need  procedures, such as the SROC curves, to make predictions. \cite{Lian-eta-al-2019-jasa} extended the model of \cite{Ma-etal-2018-biostatistics} in a SROC framework  by expressing the latent sensitivities and specificities as functions of the latent cutoff and and accuracy values. The latter latent variables are assumed to independently follow the MVN distribution.    Nevertheless,    the SROC curves can essentially show the effects of different model  assumptions, such as the choice of parametric bivariate copula and its tail dependence properties, because they are inferences that depend on the joint distribution. As the  methods in \cite{Lian-eta-al-2019-jasa} assume that the between-studies models are the MVN, the vine copula distributions will provide distributional improvements when adapted to this setting.

The model incorporates studies reporting one or more tests, but not necessarily all $K+1$ tests, under a MAR assumption, without the need to impute data on the missing binary data with debatable imputation methods (\citealt{Nyaga-etal-2018n}; \citealt{Nyaga-etal-2018b}).  The imposed dependence structure allows borrowing information between the latent disease prevalence,  sensitivities and specificities across studies.  This is emphasized for likelihood methodologies to analyse dependent data for ignorable missing-data mechanism  in \cite{Little&Rubin-2002}. By including both fixed and random effects, the dependence  structure allows for the appropriate adjustments to the parameters even when the data are incomplete \citep{BeunckensMolenberghsKenward-2005}.

Furthermore, as in this article, similarly with \cite{Ma-etal-2018-biostatistics} and \cite{Lian-eta-al-2019-jasa},  we did not consider the situation where the  tests may be conditional dependent given the latent disease status, e.g.,  
two or more index tests applied to the same patient are conditionally dependent due to a factor other than disease status, such as a biological mechanism \citep{Vacek-1985-Biometrics}. 
Further research is needed for extensions of the proposed model under such conditional dependence.

\section*{Software}
\proglang{R} functions to derive estimates and simulate from the  1-truncated C-vine copula mixed model for network meta-analysis of multiple diagnostic tests  will  be  part of the \proglang{R} package \pkg{CopulaREMADA} 
\citep{Nikoloulopoulos-2018-CopulaREMADA}.

\section*{Acknowledgements}
 The simulations presented in this paper were carried out on the High Performance Computing Cluster supported by the Research and Specialist Computing Support service at the University of East Anglia.


\end{document}